\documentclass[journal,onecolumn]{IEEEtran}
\IEEEoverridecommandlockouts
\usepackage{cite}
\usepackage{amsmath,amssymb,amsfonts}
\usepackage{algorithmic}
\usepackage{graphicx}
\usepackage{textcomp}
\usepackage{xcolor}
\usepackage{url}
\usepackage{enumerate}
\usepackage{setspace}
\usepackage{bm}
\usepackage{placeins}

\usepackage{subfig}
\usepackage{float}
\usepackage{hhline}
\usepackage{adjustbox}
\usepackage{multicol}
\usepackage{multirow}
\usepackage{makecell}

\def\BibTeX{{\rm B\kern-.05em{\sc i\kern-.025em b}\kern-.08em
    T\kern-.1667em\lower.7ex\hbox{E}\kern-.125emX}}

\newcommand  {\rr}{\ensuremath{\vec{r}}}
\newcommand  {\uu}{\ensuremath{\vec{u}}}
\newcommand  {\xx}{\ensuremath{\vec{x}}}
\newcommand  {\xxx}{\ensuremath{\vec{\bm{x}}}}
\newcommand  {\xxxx}{\ensuremath{\text{\bf x}}}

\newcommand  {\yyy}{\ensuremath{\vec{\bm{y}}}}
\newcommand  {\vv}{\ensuremath{\vec{v}}}
\newcommand  {\vvv}{\ensuremath{\vec{\bm{v}}}}
\renewcommand{\v}[1]{\ensuremath{\vec{#1}}}
\renewcommand{\u}[1]{\ensuremath{\hat{#1}}}
\newcommand {\ff}[2]{\ensuremath{\frac{#1}{#2}}}

\newcommand{\RR}{\ensuremath{{\mathbb R}}}
\newcommand{\ZZ}{\ensuremath{{\mathbb Z}}}

\begin{document}

\title{High Performance Low Complexity Multitarget Tracking Filter for a  Array of Non-directional Sensors}

\author{\IEEEauthorblockN{Christopher Thron,Khoi Tran}\\
\IEEEauthorblockA{\textit{Department of Science and Mathematics} \\
\textit{Texas A \& M University-Central Texas}\\
Killeen, TX USA \\
thron@tamuct.edu}\\
~\\
\IEEEauthorblockN{Joseph Raquepas}\\
\IEEEauthorblockA{\textit{Air Force Laboratory Information Directorate} \\
Rome, NY \\
 }}

\maketitle

\begin{abstract}
This paper develops an accurate, efficient filter (called the `TT filter')  for tracking multiple targets using a spatially-distributed network of  amplitude sensors that estimate distance but not direction. Several innovations are included in the algorithm that increase accuracy and reduce complexity. For initial target acquisition once tracking begins, a constrained Hessian search is used to find the maximum likelihood (ML) target vector, based on the measurement model and a Gaussian approximation of the prior. The Hessian at the ML vector is used to give an initial approximation of the negative log likelihood for the target vector distribution: corrections are applied if the Hessian is not positive definite due to the near-far problem. Further corrections are made by applying a transformation that matches the known nonlinearity introduced by distance-only sensors. A set of integration points is constructed using this information, which are used to estimate the mean and moments of the target vector distribution. Results show that the TT filter gives superior accuracy and lower complexity than previous alternatives such as Kalman-based or particle filters. 

\end{abstract}

\section{Introduction}
Sensor networks are becoming increasingly important in both military and civilian applications\cite{liu2007multitarget,sandell2008distributed}.
Multi-target multi-sensor tracking  in high-dimensional state spaces is still a challenging computational problem in term of time and resources. 
When a sensor network attempts to estimate the location of a set of targets at time $t$, there are two sources of information for the targets' positions: the sensor measurements at time $t$; and the prior distribution propagated from time $t-1$.  Classical tracking algorithms (such as the Kalman filter) assume that all probability distributions involved are Gaussian. However, non-directional sensor measurements give rise to probability distributions that are highly non-Gaussian. Past research has employed various particle filters to deal with the non-Gaussian nature of the distributions \cite{li2016particle}. However, particle filters typically require high computational complexity to achieve accurate results.
  
Our research takes a different approach. We take advantage of two facts: (1) typically an analytical expression is available for the measurement distribution; and (2) because of the uncertainty in propagation, the propagated prior is nearly Gaussian. In our tracking filter we use an analytical expression for the log  likelihood of the product of measurement distribution $\times$ Gaussian approximation of the prior, then subsequently refine the approximation by computing the mean and covariance of the product distribution numerically. We also make use of techniques to avoid convergence to a non-global local minimum.

In the following sections we first describe the system model used to develop and test the algorithm; we then outline  the steps of the algorithm

\section{System model description}
The system model used was taken verbatim from Li and Coates \cite{li2016particle} (and their Matlab code was used for model testing \cite{liMatlab2017}).
The system consists of a rectangular array of $S=25$ sensors with a grid spacing of 10 m. Within the region are $C=4$ targets moving stochastically (and independently) as governed by the equations:
\begin{equation}
{\xxxx}_{c,t} := F_{prop}\xxxx_{c,t-1} + \bm{\nu}_{c,t}\qquad c= 1 \ldots C; ~ t = 1 \ldots T,
\end{equation}
Where    $ {\xxxx}_{c,t} := [\xx_{c,t},\dot{\xx}_{c,t}]$ is the position+velocity vector of target $c$ at discrete time index $t$;
$F_{prop}$ is a transition matrix given by:
\begin{equation}\label{eq:Fprop}
F_{prop} := \begin{bmatrix}
1 & 0 & 1 & 0 \\
0 & 1 & 0 &1 \\
0 & 0 & 1 & 0 \\
0 & 0 & 0 & 1
\end{bmatrix},
\end{equation}

 and $\bm{\nu}_{c,t} \sim N(0,V)$ is process noise with covariance matrix $V$ where
\begin{equation}
V :=  \sigma_w^2\begin{bmatrix}
1/3 & 0 & 0.5 & 0 \\
0 & 1/3 & 0 & 0.5 \\
0.5 & 0 & 1 & 0 \\
0 & 0.5 & 0 & 1
\end{bmatrix}, \quad (\sigma_w^2 = 0.05).
\end{equation}
In the following we will often drop the $t$ index for brevity.
The expected measurement signal at sensor $s$ due to target $c$ is distance-dependent, and is  specified in terms of the quantity 
\begin{equation}\label{eq:fsc}
f_{s,c} :=  \frac{A}{||\xx_c - \xx_{[s]}||^p + d_0}, \qquad c= 1 \ldots C, s = 1 \ldots S,
\end{equation}
where $\xx_{[s]} = (x_{[s]},y_{[s]})$ is the $(x,y)$ position of sensor $s$, and $p=1$ was used. The quantity $f_{s,c}$ expresses the expected signal at sensor $s$ due to sensor $c$.  Altogether the expected signal at sensor $s$ given target positions at $\xxx :=  [ \xx_1, \ldots \xx_C]$ is:
\begin{equation}\label{eq:Fs}
\alpha_s := E[a_s |  \xxx] = \sum_{c=1}^C f_{s,c} ~~ (s=1 \ldots S).
\end{equation}
The measurement signal $a_s$ at sensor $s$ is assumed to have variance $\sigma_s^2$.

The motion process covariance is not precisely known by the observer, and the filter assumes the following covariance:
\begin{equation}\label{eq:varEst}
V' := \begin{bmatrix}
3 & 0 & 0.1 & 0 \\
0 & 3 & 0 & 0.1 \\
0.1 & 0 & 0.03 & 0 \\
0 & 0.1 & 0 & 0.03
\end{bmatrix}.
\end{equation}

\section{Algorithm description}
The steps in our multi-target estimation procedure may be outlined as follows:

\begin{enumerate}[(A)]
\item
Analytically compute the negative log likelihoods (NLL) for the measurement distribution and a Gaussian approximation to the propagated prior together with  their gradients and Hessians; 
\item
Use the analytical expressions and Hessian minimization of the  combined NLL to estimate the maximum likelihood (ML) target vector;
\item
If the NLL evaluated at the estimated target vector is above a threshold, then re-estimate the ML target vector;
\item
If the Hessian of the NLL at the ML vector is not positive definite, then perform modifications to restore positive definiteness;
\item
Generate integration points, using the inverse of the Hessian as a preliminary covariance estimate and making a nonlinear transformation for targets that are close to sensors; 
\item
Use the integration points to estimate the mean and covariance for a Gaussian approximation of the new prior;
\end{enumerate}

We fill in the mathematical details for steps (A-F) in the following subsections.

\subsection{Calculation of negative log likelihoods, gradients, and Hessians}
\subsubsection{Negative log likelihood of measurement distribution}\label{sec:NLLmeas}
We use $a_s$  to denote the received signal at sensor $s$ at time step $t$.
Based on \eqref{eq:fsc}, the NLL for the measurement distribution ${\mathcal N}^{(meas)}(\xxx)$ is given by
\begin{equation}\label{eq:nu}
{\mathcal N}^{(meas)}(\xxx)= \sum_{s=1}^S \frac{(\alpha_s - a_s)^2}{2\sigma_s^2}
\end{equation} 
To find the optimum value (minimum) of the total NLL,  we must find the gradient of ${\mathcal N}^{(meas)}(\xxx)$ with respect to $[ \xx_1, \ldots \xx_C]$ . Based on 
\eqref{eq:Fs}-\eqref{eq:nu}, the gradient with respect to $\xx_c$ may be given in terms of the gradients of $f_{s,c}$:
\begin{equation}\label{eq:gradNmeas}
\vec{\nabla}_{\xx_c} {\mathcal N}^{(meas)}(\xxx)= \sum_{s=1}^S \sum_{c=1}^C  \left( \frac{\alpha_s - a_s}{\sigma_s^2} \right) \vec{\nabla}_{\vec{x}_c} f_{s,c}.
\end{equation}
Note that $\vec{\nabla}_{\vec{x}_c} f_{s,c} = \vec{\nabla}_{\rr_{c,s}} f_{s,c}$ where $\rr_{c,s} \equiv \xx_c - \xx_{[s]}$. It follows that
\begin{align}
\vec{\nabla}_{\vec{x}_c} f_{s,c} &= \frac{-(p/2)A(||\rr_{c,s}||^2)^{p/2-1}  }{(||\rr_{c,s}||^p + d_0)^2} \rr_{c,s} \\
& = \frac{-(p/2)A(||\rr_{c,s}||^2)^{p/2-1}  }{(||\rr_{c,s}||^p + d_0)^2} \rr_{c,s} \\
& \equiv g(||\rr_{c,s}||^2) \rr_{c,s} 
\end{align}
By taking another derivative, we may find the Hessian $H {\mathcal N}^{(meas)}$:
\begin{equation}\label{eq:HNmeas}
\begin{aligned}
H {\mathcal N}^{(meas)}(\xxx)= &\sum_{s=1}^S \sum_{c=1}^C  \left( \frac{1}{\sigma_s^2} \right) (\vec{\nabla}_{\vec{x}_c} f_{s,c})  \vec{\nabla}_{\vec{x}_c} f_{s,c}^T 
+ \sum_{s=1}^S \sum_{c=1}^C  \left( \frac{\alpha_s - a_s}{\sigma_s^2} \right)   \vec{\nabla}_{\vec{x}_c} ( \vec{\nabla}_{\vec{x}_c}  f_{s,c}),
\end{aligned}
\end{equation}
where (writing $g(||\rr_{c,s}||^2)$ as $g$ for brevity)
\begin{equation}
\begin{aligned}
\vec{\nabla}_{\vec{x}_c}( \vec{\nabla}_{\vec{x}_c} f_{s,c} ) &= \vec{\nabla}_{\vec{x}_c}( g \cdot \rr_{c,s} ) \\
& =  \frac{dg~~~~~~~}{d (||\rr_{c,s}||^2)} (2 \rr_{c,s} \rr_{c,s}^{\,T}) + g \cdot I \\
&= 2 g \cdot \left[ \left(\frac{p/2 - 1}{||\rr_{c,s}||^2} - \frac{2}{||\rr_{c,s}||^p + d_0} \right)(\rr_{c,s}^{\,T} \rr_{c,s} ) +   I \right] \\
\end{aligned}
\end{equation}

\subsubsection{NLL of propagated prior distribution}
The analytical expression for the NLL of the propagated prior may be computed as follows. 
We assume that the prior distribution is  joint Gaussian with mean $(\v{m}_{\xxx,t-1} ; \v{m}_{\vvv,t-1})$ and
covariance $\Sigma_{t-1}$, so that
\begin{equation}
p_{t-1}(\xxx,\vvv)  \propto  \exp \left(-(\xxx - \v{m}_{\xxx,t-1};\vvv - \v{m}_{\vvv,t-1})^T    \cdot \Sigma_{t-1}^{-1} (\xxx - \v{m}_{\xxx,t-1};\vvv - \v{m}_{\vvv,t-1}) / 2 \right),
\end{equation}
$(\v{m}_{\xxx,t-1} ; \v{m}_{\vvv,t-1})$ and $\Sigma_{t-1}$ are the prior's mean vector and covariance matrix, respectively.
This distribution is propagated according to the motion model \eqref{eq:Fprop}, except that the actual process noise covariance is replaced with the estimated covariance $V'$ defined in \eqref{eq:varEst}. It follows that the propagated spatial distribution from time $t-1$  (i.e., the prior distribution at time $t$) is also Gaussian with mean
\begin{equation}\label{eq:propMean} 
(\v{m}^{(prop)}_{\xxx} ; \v{m}^{(prop)}_{\vvv}) \equiv  F_{prop} (\v{m}_{\xxx,t-1}; \v{m}_{\vvv,t-1})
\end{equation}
 and covariance 
\begin{equation}\label{eq:propCov}
\Sigma_{prop} \equiv F_{prop} \Sigma_{t-1} F_{prop}^T + V'.
\end{equation}
  It follows that the NLL of propagated distribution is ${\mathcal N}^{(prop)}$, where
\begin{equation}\label{eq:Nprop}
{\mathcal N}^{(prop)}(\xxx) = \frac{1}{2}(\xxx - \v{m}^{(prop)}_{\xxx})^T  \Sigma_{prop,\xxx \xxx}^{-1}(\xxx - \v{m}^{(prop)}_{\xxx})
\end{equation}
The gradient is 
\begin{equation}\label{eq:gradNprop}
\vec{\nabla} {\mathcal N}^{(prop)}(\xxx) = \Sigma_{prop,\xxx \xxx}^{-1}(\xxx - \v{m}^{(prop)}_{\xx})
\end{equation}
and
\begin{equation}\label{eq:HNprop}
H {\mathcal N}^{(prop)}(\xxx) = \Sigma_{prop,\xxx \xxx}^{-1}.
\end{equation}

\subsection{Estimation of maximum likelihood target vector (spatial components)}
The minimizer of ${\mathcal N} \equiv {\mathcal N}^{(meas)} + {\mathcal N}^{(prop)}$ corresponds to the ML estimator of the joint location vector for the targets. The Matlab routine ``fmincon'' was used for the minimization. Constraints were added to require the targets to remain within the grid region. In the minimization, the analytical gradient and Hessian were used, where:
\begin{equation}
\vec{\nabla}{\mathcal N}(\xxx)  = \vec{\nabla}{\mathcal N}^{(meas)}(\xxx) + \vec{\nabla}{\mathcal N}^{(prop)}(\xxx)  \text{ and }
H{\mathcal N}(\xxx)  = H{\mathcal N}^{(meas)}(\xxx) + H{\mathcal N}^{(prop)}(\xxx),
\end{equation}
and $\vec{\nabla}{\mathcal N}^{(meas)}$, $\vec{\nabla}{\mathcal N}^{(prop)}$, $H{\mathcal N}^{(meas)}$, and $H{\mathcal N}^{(prop)}$ are given by 
\eqref{eq:gradNmeas}, \eqref{eq:gradNprop}, \eqref{eq:HNmeas}, and \eqref{eq:HNprop} respectively.

\subsection{Consistency check and re-estimation of the target vector}\label{sec:consistency}
It is possible that the algorithm may lose track of the targets, or may fail in initial acquisition. Thus we introduce a check for whether or not the ML target location vector is consistent with the measurement distribution. In this section the checking procedure is described, as well as the re-estimation procedure that is applied if the threshold is not met.

According to the signal model, the signal $a_s$ at sensor $s$ ($s=1\ldots S$) is Gaussian distributed with mean $\alpha_s$ and variance $\sigma_s$, and the signals at different sensors are independent. It follows that the the sum of squared normalized signal deviations given by ${\mathcal N}^{(meas)}(\xxx)$ in \eqref{eq:nu} has a Chi-squared distribution with $S$ degrees of freedom.  This statistic may be computed using signal data and the ML target position vector, and consistency is rejected if the signal exceeds a threshold computed using the known distribution of $\chi^2_S$ and a user-specified $p$-value (in our simulations, we used $p = 0.0013$, corresponding to 3 standard deviations). 

In case the signal does not meet the threshold, a correction procedure is applied to find a target position vector with higher likelihood.  Two different correction procedures were developed, as described below.

\subsubsection{One-by-one sensor addition correction procedure}\label{sec:1by1}
The procedure is based on the observation that when signals from too many sensors are combined, the resulting function has many suboptimal local minima (during optimization, it is easy for the target estimate to get caught in the middle of one of the squares in the grid formed by the sensors). On the other hand, when some sensors are excluded and only sensors that are far from the targets are used, then local minima are avoided but the accuracy of the target position estimates is reduced. So we begin with far-away sensors to steer the target vector to a neighborhood of the correct solution. We then add sensor signals one by one to gradually increase the accuracy without introducing nearby local minima that could prevent the solution from reaching the global optimum.  The steps in the procedure may be described as follows:
\begin{enumerate}[\hspace{1cm}(a)]
\item
Estimate the target vector, using only sensors on the boundary of the region;
\item
Loop over remaining (unused) sensors:
\item
Identify the unused sensor which maximizes the minimum distance to current estimated target positions;
\item
Add this ``maximin'' sensor to the list of sensors used to estimate the target vector, and re-estimate the target vector;
\item
end loop
\end{enumerate}
\subsubsection{Square-hopping  correction procedure}\label{sec:hopping}
The second correction procedure is somewhat more complicated than the first. This procedure first locates estimated target positions that should  excessively large signals at some sensors (compared to the measured signals), then moves these targets near to sensors that are receiving larger signals. For this purpose, we define the ``signal excess'' due to target $c$ as $\epsilon(c)$ where
\begin{equation}
\epsilon(c) = \sum_{s=1}^S \max( (\alpha_s -  a_s) - Af_{s,c} , 0). 
\end{equation}
$\epsilon(c)$ can be interpreted as the sum over sensors of the reduction in (predicted$-$observed) signal that is achieved when the signal from target $c$ is removed. A large value of $\epsilon(c)$ is an indicator that target $c$ is wrongly placed.

On the other hand, we may define the ``signal deficit'' at sensor $s$ as

Using these definitions, we may outline the ``square-hopping'' procedure as follows:
\begin{enumerate}[\hspace{1cm}(a)]
\item
Identify the set of $n_{bad\_tgt}$ of targets associated with the largest values of $\epsilon(c)$  (our algorithm used $n_{bad\_tgt}=2$);
\item
Remove the targets identified in step (a) and recalculate the expected signal at each sensor
\item
Locate the $n_{bad\_sq}$ grid squares where the sensors at the corners have the largest combined signal deficit
(our algorithm used $n_{bad\_sq} = 12$);
\item
List the possible subsets of size $n_{bad\_tgt}$ of the grid squares computed in step (c);
\item
For each subset identified in step (d), place a target at the middle of each square and then run the minimizing algorithm to estimate the MLE target vector
\item
If the NLL of the target vector identified in step (e) is below the NLL acceptance threshold, then use the resulting target vector as MLE
\item
If the NLL of the target vector identified in step (e) is not below the NLL  acceptance threshold, then return to step (d)
end loop
\end{enumerate}

\subsection{Hessian correction}
Although theoretically the Hessian should be positive definite at a function minimum, in practice the Hessian may fail to be positive definite because of numerical issues.  This can happen when at least one target is very close to a sensor.  In this case,  the target in question is identified by means of the measurement signal, and rows and columns of the Hessian corresponding to that particular target are modified. 
A description of the Hessian correction procedure is as follows:
\begin{enumerate}[\hspace{1cm}(a)]
\item
Initialize an empty list of excluded target-sensor pairs
\item
Using the current target vector estimate, locate the closest target-sensor pair that is not in the list, and add to the list
\item 
Find the MLE target vector estimate with signals from sensors in the excluded list removed, and with targets in the excluded list
fixed at their current positions (so they are not free parameters in the optimization)
\item
If the Hessian for the non-excluded targets is not positive definite, then go back to step (b)
\item
If the Hessian for the non-excluded targets is positive definite, add rows and columns to the Hessian corresponding to the excluded targets.
For added rows and columns, all non-diagonal entries are 0, and the diagonal entry is the fixed value $d_0^{-2}$, which corresponds to a target position variance of $d_0^2$ (note the inverse of the Hessian is the
estimated covariance for the measurement distribution).  
\end{enumerate}

\subsection{Numerical integration of the combined distribution}
\subsubsection{Integral with Gaussian weight function}
Using the method described above, by minimizing ${\mathcal N}(\xxx)$  we obtain an estimate $\v{m}_{\xxx}$ for the mean of the target vector's spatial distribution at time $t$ (denoted by $p(\xxx)$).  We may also use the Hessian $H \equiv H{\mathcal N}(\v{m}_{\xxx})$ as a preliminary estimate of the inverse of the spatial covariance. At this point, our goal is to compute better estimates of the spatial mean and covariance via numerical integration. We suppose that the distribution is approximately Gaussian with mean   $\v{m}_{\xxx}$ and spatial covariance $H^{-1}$. 
In order to compute moments of the distribution, we will need to compute integrals of the form:
$\int_{\RR^d}  f(\xxx) p(\xxx) d^d \xxx.$ We may change variable as follows.
Since $H$ is the Hessian at a function's minimum, it is a positive definite symmetric matrix and thus has a Cholesky square root $H^{1/2}$. 
We may define:
\begin{equation}\label{eq:ydef}
\yyy(\xxx) \equiv H^{1/2}(\xxx - \v{m}_{\xxx}) 
\implies \xxx(\yyy) = H^{-1/2}\yyy + \v{m}_{\xxx}.
\end{equation}

The actual distribution is modeled as a perturbation of the Gaussian distribution with mean $\v{m}_{\xxx}$ and spatial covariance $H^{-1}$, so the distribution function can be expressed as:

\begin{equation}\label{eq:pme} 
p (\xxx) \approx C g(\yyy(\xxx)) \exp \left( -\ff{1}{2} \yyy^T \yyy \right),
 \end{equation}
so that 
\begin{equation}\label{eq:pme2} 
g (\yyy(\xxx)) \approx C^{-1} p(\xxx) \exp \left( \ff{1}{2} \yyy^T \yyy \right),
 \end{equation}
where $g(\yyy)$ is the perturbation function and  $C$ is a normalizing constant.

This gives us
\begin{equation}
\int_{\RR^d}  f(\xxx) p(\xxx) d^d \xxx   =  \sqrt{|H|} \int_{\RR^d} h(\yyy) \exp \left( -\ff{1}{2} \yyy^T \yyy \right) d^d \yyy
\end{equation}
where 
\begin{equation}\label{eq:h}
h(\yyy) \equiv  C f\left(\xxx(\yyy)\right) g(\yyy).
\end{equation} 
We may write the $\yyy$ integral  in polar coordinates as:
\begin{equation}
\int_0^{\infty} \left( \int_{S^7}  h(r, \Theta)    d \Theta \right)  r^{d-1} \exp \left( -\ff{r^2}{2}\right)  dr,
\end{equation}
and changing variable $s \equiv r^2/2$ gives:
\begin{equation}\label{eq:GL_int1}
2^{d/2-1} \int_0^{\infty} \left( \int h(\sqrt{2s}, \Theta)d \Theta \right) s^{d/2-1} e^{-s}  ds,
\end{equation}
\subsubsection{Application of Gauss-Laguerre quadrature}
The  $s$ integral can be approximated using Gauss-Laguerre quadrature\cite{laguerreQuad}. $n$-point Gauss-Laguerre quadrature depends on the generalized Laguerre polynomials $L^{(d/2-1)}_{n}(z)$ and $L^{(d/2)}_{n+1}(z)$, and will integrate exactly polynomials in $s$ of order $2n-1$.  From the recurrence relation in \cite{laguerrePoly}, we find for the case $d=8$:
\begin{align}
L^{(3)}_0(z) &= 1 \\
L^{(3)}_1(z) &= 1 + 3 -z =4-z\\
L^{(3)}_2(z) &= \ff{(3+3-z)L^{(3)}_1(z) - (1+3)L^{(3)}_0(z)}{2} \\
 & = \ff{20 - 10x + z^2}{2}\\
L^{(3)}_3(z) &= \ff{(5+3-z)L^{(3)}_2(z) - (2+3)L^{(3)}_1(z)}{3} \\
& = \ff{8-z}{3}L^{(3)}_2(z) - \frac{5}{3}(4 - z)
\end{align}
For two-point Gauss-Laguerre quadrature, the quadrature points are the roots of $L^{(3)}_2(z)$, which are $z_\pm = 5 \pm \sqrt{5}$. The weights are
\begin{equation}
w_\pm = \ff{(2+3)! \,z_\pm}{2!(2+1)^2 (L_3^{(3)}(z_\pm))^2},
\end{equation}
and noting that  $L_3^{(3)}(x_\pm) = - \ff{5}{3}(4 - x_\pm)$, we have
\begin{equation}
w_\pm = \left(\ff{6}{5}\right) \ff{5 \pm \sqrt{5}}{3 \pm \sqrt{5}}.
\end{equation}
It follows that the integral in \eqref{eq:GL_int1} can be approximated as:
\begin{equation}
2^{d/2-1} \left( w_{-} \int h(\sqrt{2z_{-}}, \Theta)d \Theta   + w_{+} \int h(\sqrt{2z_{+}}, \Theta)d \Theta \right).
\end{equation} 
\subsubsection{Performing angular integration using zero-sum lattice points}
Supposing that we are in $d$-dimensional space ($d=8$ in our particular case), we need to perform the angular integral over a $(d-1)$-dimensional sphere. This angular integral can be approximated by a sum over the mesh points of a mesh that uniformly covers the sphere. One choice of mesh is the $d(d+1)$ unit vectors in the root system of the $d$-dimesional simplex (zero-sum) lattice. 
The zero-sum lattice may be defined as the sublattice of $\ZZ^{d+1}$ restricted to the plane  $x_1 + \ldots + x_d +x_{d+1}= 0$\cite{bewley2011new}. To obtain a lattice in $\RR^d$, we need to rotate this plane into the plane $x_{d+1}=0$. The easiest way to do this is to rotate the unit vector $\v{1}$ in $\RR^{d+1}$ so that it aligns along the elementary basis vector $\u{e}_{d+1}$, while leaving all orthogonal vectors fixed. We denote the desired rotation by $\rho$. Let $\mathcal P$ denote the plane determined by the two vectors $\v{1}$ and $\u{e}_{d+1}$: note that $\mathcal P$ is fixed by $\rho$, and all vectors orthogonal to $\mathcal P$ are invariant under $\rho$. Let $\u{\alpha} \equiv \frac{\v{1} - \u{e}_{d+1}}{\sqrt{d}}$, so that $\mathcal P$ is spanned by the orthonormal set$\{\u{\alpha}, \u{e}_{d+1}\}$ Then if $1 \le j \le d$, we may express $\u{e}_j$ as the sum of two orthogonal vectors:
\begin{equation}
\u{e}_j = (\u{e}_j \cdot \u{\alpha})\u{\alpha} + (\u{e}_j - (\u{e}_j \cdot \u{\alpha})\u{\alpha}).
\end{equation}
Since  $\u{e}_j \cdot \u{\alpha} = \frac{1}{\sqrt{d}}$ for all $j \le d$, we have:
\begin{equation}
\rho(\u{e}_j) =   \frac{\rho(\u{\alpha})}{\sqrt{d}}   + \rho \left(\u{e}_j - \frac{\u{\alpha}}{\sqrt{d}}\right) .
\end{equation}  
But the vector $\u{e}_j - \frac{\u{\alpha}}{\sqrt{d}}$ is orthogonal to $\mathcal P$, and is thus fixed under $\rho$.  This implies:
\begin{equation}\label{rho_ej}
\rho(\u{e}_j) =   \u{e}_j  + \frac{\rho(\u{\alpha}) - \u{\alpha}}{\sqrt{d}}.
\end{equation}
The set of closest neighbors of $\v{0}$ in the restricted zero-sum lattice in $\RR^{d+1}$ consists of all vectors of the form $\u{e}_i - \u{e}_j$ where $1 \le i,j \le d+1$.  When rotating this set into $\RR^d$ we only need to compute $\rho(\u{e}_j - \u{e}_{d+1})$, where $1 \le j \le d$, since all other vectors in the set are invariant under $\rho$.   From \eqref{rho_ej} we have 
\begin{equation}
\rho(\u{e}_j - \u{e}_{d+1}) =  \u{e}_j  + \v{q},
\end{equation}
where
 \begin{equation}
 \v{q} \equiv   \frac{\rho(\u{\alpha} - \u{e}_{d+1}) - \u{\alpha}}{\sqrt{d}}
 \end{equation}
 Now we have that $\v{q} \in \mathcal P$ and $\v{q} \perp \u{e}_{d+1}$, which implies that $\v{q} \parallel \u{\alpha}$. 
The additional condition that $|| \u{e}_j  + \v{q} ||^2=2$ implies that.
\begin{equation}\label{eq:q}
\v{q} = \ff{-1 \pm \sqrt{d+1}}{d} (\underbrace{1,1,\ldots , 1}_{d~\textrm{terms}},0).
\end{equation} 
There are two possible choices for $\v{q}$ (corresponding to rotating $\v{1}$ parallel or antiparallel to $\u{e}_{d+1}$), and either may be chosen.   Since the final entry of $\v{q}$ in \eqref{eq:q} is 0, it follows $\v{q} \in \RR^d$, and we may drop the final `0'. 
Consequently, we may generate the $d(d+1)$ vectors in  the root system of the simplex lattice in $\RR^d$ as follows:
 \begin{itemize}
 \item
 Take all ordered pairs $(i,j)$ where $1 \le i,j \le d$~~  ($d(d-1)$ vectors), and form the vectors $\u{e}_i - \u{e}_j$;
 \item 
 for $1 \le j \le d$, form the vectors $\u{e}_j + \v{q}$ and $-\u{e}_j - \v{q}$.
 \end{itemize}
 
% The root system should be rotated so as to cause mixture in the coordinates.  We may do this by successively rotating subspaces.
% Subspace rotation may be implemented as follows (when $d$ is even:
% 
% \begin{itemize}
%     \item Generate the $d \times d$ complex DFT matrix;
%     \item Select every other column, beginning with the second column (resulting in 
%     a $d \times d/2$ matrix);
%     \item Split each column into two columns (real and imaginary parts);
%     \item Replace each single column with the two columns obtained in the previous step, resulting in a $d \times d$ matrix.
% \end{itemize}
% 
 %\noindent
 %Given an orthonormal basis $\{\u{b}_1, \ldots, \u{b}_n \}$ in columns (input as a matrix):
 % \begin{itemize}
 %\item Compute the vector $\v{c}_1\equiv \u{b}_1 + 2\u{b}_2 + \cdots + n\u{b}_n$;
 %\item normalize the vector to obtain $\u{c}_1$
 %\item Apply Gram-Schmidt orthogonalization to the basis $\{\u{c}_1, \u{b}_2, \ldots, \u{b}_n \}$ to obtain an orthonormal basis
%  $\{\u{c}_1, \u{c}_2, \ldots, \u{c}_n \}$ (output the basis as a matrix with the  $\u{c}_j$'s as column vectors.
%  \end{itemize}
%  The new orthonormal basis is $\{\u{c}_1, \ldots, \u{c}_n \}$.  The algorithm is called with identity matrix $I$ as input, and may be repeated with the $n-1$ column vectors $\{\u{c}_2, \ldots, \u{c}_{n} \}$,   and subsequently iterated. The final result is a rotation matrix that is applied to the  root system vectors.  
   
We may determine weights on the grid points $\{\v{\Theta}_j, j = 1,\ldots J\}$ where $J \equiv d(d+1)$ as follows. The area of a sphere in $d$ dimensions (when $d$ is even) is $\frac{2\pi^{d/2}}{(d/2-1)!}$. It follows that $\int d\Theta = \frac{2\pi^{d/2}}{(d/2-1)!}$. In order to give the correct area integral, then each point in the mesh should be assigned a weight   $\frac{2\pi^{d/2}}{(d/2-1)!(d)(d+1)}$. It follows that the integral in \eqref{eq:GL_int1} can be approximated as:
\begin{equation}\label{eq:GL_int2}
 A \sum_{j=1}^J \left( w_{-} h(\sqrt{2z_{-}}\v{\Theta}_j)   + w_{+}   h(\sqrt{2z_{+}}\v{\Theta}_j) \right),
\end{equation} 
where
\begin{equation}
A\equiv \frac{(2\pi)^{d/2}}{(d/2-1)!(d)(d+1)}.
\end{equation}
\subsubsection{Numerical integration summary}
Based on \eqref{eq:GL_int2}, \eqref{eq:pme} and \eqref{eq:h} the integral $\int_{\RR^d}  f(\xxx) p(\xxx) d^d \xxx $ can be approximated as:
\begin{equation}\label{eq:int2}
\begin{aligned}
\int_{\RR^d}  f(\xxx) p(\xxx) d^d \xxx 
& \approx  
\sqrt{|H|} AC \sum_{j=1}^{J} \left( w_{-} h(\sqrt{2z_{-}}\v{\Theta}_j)   + w_{+}   h(\sqrt{2z_{+}}\v{\Theta}_j) \right),\\
& \approx  
\sqrt{|H|}  AC \sum_{j=1}^{J}  w_{-} e^{z_{-}} f(\xx_{j}^{\,-}) p(\xx_{j}^{\,-}) 
+ w_{+} e^{z_{+}}  f(\xxx_{j}^{\,+}) p(\xxx_{j}^{\,+}) 
\end{aligned}
\end{equation}
where
\begin{equation}\label{eq:ydef2}
\xxx_{j}^{\,\pm} \equiv \sqrt{2z_{\pm}}H^{-1/2}\v{\Theta}_j + \v{m}_{\xxx}.
\end{equation}
The relation between p(\xxx) and ${\mathcal N}(\xxx)$ is
\begin{equation}
p(\xxx) = B \exp(-{\mathcal N}(\xxx)).
\end{equation}
We may avoid calculating normalizing constants by recognizing that $\int_{\RR^d}  f(\xxx) p(\xxx) d^d \xxx = 1$, which leads to:
\begin{equation}\label{eq:int}
\begin{aligned}
1 &= \sqrt{|H|}  ABC \sum_{j=1}^{J} \left( w_{-} e^{z_{-}}\exp(-{\mathcal N}(\xxx_{j}^{\,-})  + w_{+} e^{z_{+}}  \exp(-{\mathcal N}(\xxx_{j}^{\,+}) \right)
\end{aligned}
\end{equation}
It follows that 
\begin{equation}\label{eq:intFinal}
\begin{aligned}
&\int_{\RR^d}  f(\xxx) p(\xxx) d^d \xxx   \approx  
C^{\prime} \sum_{j=1}^{J}  \left(w_{-} e^{z_{-}} f(\xxx_{j}^{\,-}) \exp(-{\mathcal N}(\xxx_{j}^{\,-})   +  w_{+} e^{z_{+}}  f(\xxx_{j}^{\,+}) \exp(-{\mathcal N}(\xxx_{j}^{\,+}) \right).
\end{aligned}
\end{equation}
where
\begin{equation}
(C^{\prime})^{-1} \equiv \sum_{j=1}^{J}  w_{-} e^{z_{-}}\exp(-{\mathcal N}(\xx_{j}^{\,-}))   
+  w_{+} e^{z_{+}}  \exp(-{\mathcal N}(\xxx_{j}^{\,+}) ).
\end{equation}

Summarizing, we have:
\begin{equation}\label{eq:intFinal1}
\int_{\RR^d}  f(\xxx) p(\xxx) d^d \xxx   \approx  
C^{\prime}  \sum_{k=1}^{2J}  p_k f(\xxx_{k}), 
\end{equation}
where
\begin{equation}\label{eq:defFinal}
\begin{aligned}
\xx_{k}&\equiv
\begin{cases}  
\sqrt{2z_{-}}H^{-1/2}\v{\Theta}_k + \v{m}_{\xxx} & 1 \le k \le J ;\\
\sqrt{2z_{+}}H^{-1/2}\v{\Theta}_{k-J} + \v{m}_{\xxx} & J+1 \le k \le 2J, 
\end{cases}
\end{aligned}
\end{equation}
\begin{equation}
\begin{aligned}
p_{k}&\equiv
\begin{cases}  
w_{-} e^{z_{-}} \exp(-{\mathcal N}(\xxx_{k})) & 1 \le k \le J; \\
w_{+} e^{z_{+}} \exp(-{\mathcal N}(\xxx_{k})) & J+1 \le k \le 2J, 
\end{cases}
\end{aligned}
\end{equation}
and
\begin{equation}
C^{\prime} \equiv \left( \sum_{j=1}^{2J} p_k \right)^{-1}.   
\end{equation}

\subsubsection{Nonlinear transformation of measurement distribution}\label{sec:nonlinear}
When a target is close to a sensor $s$, the measurement distribution is not Gaussian in Cartesian coordinates. Instead, the measurement distribution is better approximated
by a Gaussian in polar coordinates, where the origin of the coordinate system is the closest sensor to the target. The coordinates transformation is given by:
\begin{equation}\label{eq:xy2uv}
\begin{aligned}
u& \equiv \sqrt{(x - x_{[s]})^2 + (y - y_{[s]})^2)};\\
v& \equiv  u \, \text{atan2}( y - y_{[s]}, x - x_{[s]}).
\end{aligned}
\end{equation}
A Gaussian distribution in these coordinates has the form 
\begin{equation}
p_{\uu}(\uu) \equiv C^{\prime \prime} \exp( -(\uu-\v{m}_{\uu})^T \Sigma_{\uu ,\uu } (\uu-\v{m}_{\uu}) ),
\end{equation}
where
\begin{equation}  
\begin{aligned}
\uu &\equiv (u,v);\\
\v{m}_{\uu} &\equiv \sqrt{x_{c,s}^2 + y_{c,s}^2} 
\begin{bmatrix}
1 \\ \text{atan2}( y_{c,s}, x_{c,s} )
\end{bmatrix};\\
x_{c,s} &\equiv x_c - x_{[s]}; \qquad y_{c,s} \equiv y_c - y_{[s]}.
\end{aligned}
\end{equation}
It follows that expected values may be computed (in analogy to \eqref{eq:intFinal1}) as 
$ \int_{\RR^d}  f(\,\xx(\uu) \, ) p_{\uu}(\uu ) d^d \uu$
By changing variable we may write this as an integral in $\xx$:
\begin{equation}
\int_{\RR^d}  f(\xx) p_{\uu}(\,\uu (\xx)\, )   \left| \frac{\partial \uu}{\partial \xx} \right| d^d \xx,
\end{equation}
where
\begin{equation}
\begin{aligned}
\frac{\partial \uu}{\partial \xx} &= 
\frac{1}{r}\begin{bmatrix} 
x & y \\ x \, \text{atan2}(y,x) - y & y \, \text{atan2}(y,x) + x
\end{bmatrix}; \\
\left| \frac{\partial \uu}{\partial \xx} \right| &= 1.
\end{aligned}
\end{equation}

\begin{equation}\label{eq:defFinal2}
\begin{aligned}
\uu_{k}&\equiv
\begin{cases}  
\sqrt{2z_{-}}\Sigma_{\uu \uu}^{1/2}\v{\Theta}_k + \v{m}_{\uu} & 1 \le k \le J \\
\sqrt{2z_{+}}\Sigma_{\uu \uu}^{1/2}\v{\Theta}_{k-J} + \v{m}_{\uu} & J+1 \le k \le 2J 
\end{cases}
\end{aligned}
\end{equation}

The covariance in $\uu$ coordinates $\Sigma_{\uu \uu}$ may be computed using the local transformation of $\Sigma_{\xx \xx}$: 
\begin{equation}
\Sigma_{\uu \uu} = \left. \frac{\partial \uu}{\partial \xx} \right|_{\xx = \xx_{c,s}}  \cdot  \Sigma_{\xx \xx} \cdot \left. \frac{\partial \uu}{\partial \xx} \right|_{\xx = \xx_{c,s}} ^{-1},
\end{equation}

In order to transform the points $\{\uu_{k}\}$ back to $(x,y)$ coordinates, we use the inverse transformation to \eqref{eq:xy2uv}:
\begin{equation}\label{eq:uv2xy}
x_k = u_k \cos (v_k/u_k) + x_{[s]}; \qquad y_k = u_k \sin(v_k/u_k) + y_{[s]};
\end{equation}

\subsection{Updating mean and covariance}
The updated mean and spatial covariance matrix at time $t$ are given by:
\begin{equation}
\begin{aligned}
\v{m}_{\xx,t} &= \sum_{k=0}^{2J} \xx_{k} p_k \\
\Sigma_{\xx \xx, new} &= \sum_{k=0}^{2J} \xx_{k} \xx_{k}^T p_k -  \v{m}_{\xx,t} \v{m}_{\xx,t}^T.
\end{aligned}
\end{equation}
% To get a better estimate of the mean and covariance, we may replace $\v{m}_{\xx}$ and $H$ in \eqref{eq:defFinal} by $\v{m}_{\xx,new}$ and $\Sigma_{\xx\xx,new}^{-1}$ respectively, and iterate until convergence.

We must also compute the updated mean and covariance for the velocities for time $t$. We may use the following facts, which hold for Gaussian distributions\cite{wang2020}:
\begin{align}
\v{m}_{\vv | \xx} &= \v{m}_{\vv}^{(prop)} + \Sigma_{\vv \xx}^{(prop)} (\Sigma_{\xx \xx}^{(prop)})^{-1}(\xx - \v{m}_{\xx}^{(prop)}); \\
\Sigma_{\vv \vv | \xx} &= \Sigma_{\vv \vv}^{(prop)} - \Sigma_{\vv \xx}^{(prop)} (\Sigma_{\xx \xx}^{(prop)})^{-1} \Sigma_{\xx \vv}^{(prop)}.
\end{align} 
so we may write:
\begin{align}
\v{m}_{\vv,t} &= \sum_{k=1}^{2J} \v{m}_{\vv | \xx_k} p_k  =   \v{q} +  Q \v{m}_{\xx},
\end{align}
where
\begin{align}
\v{q} &\equiv \v{m}_{\vv}^{(prop)} - \Sigma_{\vv \xx}^{(prop)} (\Sigma_{\xx \xx}^{(prop)})^{-1} \v{m}_{\xx}^{(prop)} \text{ and }Q \equiv \Sigma_{\vv \xx}^{(prop)} (\Sigma_{\xx \xx}^{(prop)})^{-1}.
\end{align} 

We also have
\begin{align}
&E[\vv \vv^T \, |\, \xx] = \Sigma_{\vv \vv | \xx} + \v{m}_{\vv | \xx} \v{m}_{\vv | \xx}^T \implies E[\vv \vv^T] =   \Sigma_{\vv \vv | \xx} + \sum_{k=1}^{2J}   \v{m}_{\vv | \xx_k} \v{m}_{\vv | \xx_k}^T p_k,
\end{align}
so that
\begin{align}
E[\vv \vv^T] =  &\Sigma_{\vv \vv | \xx} +  \sum_{k=1}^{2J}   (\v{q} + Q \xx_k) (\v{q} + Q \xx_k)^T p_k\\
 =  &\Sigma_{\vv \vv | \xx} + \v{q}\v{q}^T + \v{q} (Q \v{m}_{\xx})^T + (Q \v{m}_{\xx}) \v{q}^T + \sum_{k=1}^{2J}   Q \xx_k (Q \xx_k)^T p_k.
\end{align}
This gives the following expression for the velocity covariance:
\begin{align}
&\Sigma_{\vv \vv,t } = E[\vv \vv^T] - E[\vv]E[\vv^T] \\
& \quad  =  \Sigma_{\vv \vv | \xx} + \sum_{k=1}^{2J}   Q \xx_k (Q \xx_k)^T p_k 
-  Q \v{m}_{\xx} (Q \v{m}_{\xx})^T .
\end{align}
The cross-covariance between velocity and position is also needed:
\begin{align}
\Sigma_{\vv\xx,t} &=E[\vv \xx^T] - E[\vv]E[\xx]^T \\
&= \sum_{k=1}^{2J} \v{m}_{\vv | \xx_k} \xx_k^T p_k - \v{m}_{\vv,t} \v{m}_{\xx,t}^T \\ 
&= \sum_{k}^{2J} p_k (Q \xx_k) (\xx_k -  \v{m}_{\xx,t})^T
\end{align}
The complete covariance matrix estimate  $\Sigma_{t}$ is composed of the blocks $\Sigma_{\xx\xx,t}$,  $\Sigma_{\vv\vv,t}$, $\Sigma_{\vv\xx,t}$, $\Sigma_{\vv\xx,t}^T$ as follows:
\begin{equation}
\Sigma_{t} = \begin{bmatrix}
\Sigma_{\xx\xx,t} & \Sigma_{\vv\xx,t}^T\\
 \Sigma_{\vv\xx,t} & \Sigma_{\vv\vv,t}
\end{bmatrix}
\end{equation}
The vector $(\v{m}_{\xx,t}; \v{m}_{\vv,t})$ and matrix $\Sigma_{t}$ are then used in \eqref{eq:propMean} and\eqref{eq:propCov} to obtain the propagated Gaussian prior for time $t+1$, which (when combined with the spatial distribution derived from the measurement) produces position and velocity estimates for time $t+1$. Tracking of the targets proceeds by iterating this process.

\section{Simulations}
\subsection{Simulation setup}
\begin{figure}[htb]
	\begin{center}
		\includegraphics[width=4in]{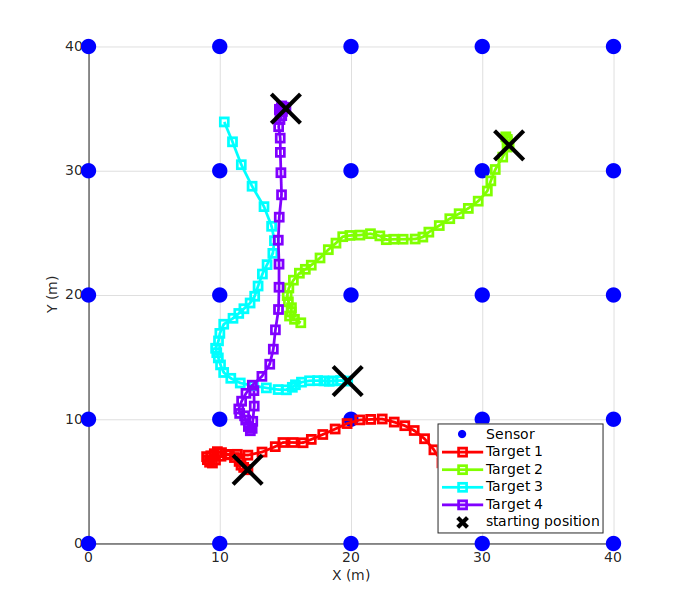}
			\caption{Average OMAT error per time step for baseline parameters}
		\label{fig:sensorNetwork}
	\end{center}
\end{figure}

The simulation was set up with 4 moving targets in an area of 40 meters by 40 meters with 25 omnidirectional sensors spaced 10 meters apart as shown in
\ref{fig:sensorNetwork}.

The mathematical model used was described in previous sections. Besides the default parameter values, the following parameter values were used:

\begin{table}[hb]
\begin{center}
 \begin{tabular}{|m{3cm} | m{7cm} | m{4cm} | } 
\hline
Parameter symbol& Significance & Values \\
 \hhline{|=|=|=|}
$\qquad \sigma_w^2$ & Variance of measurement noise &  0.0001, 0.001, \emph{0.01}, 0.1, 1 \\
$\qquad \alpha$ & Spatial covariance of filter’s process noise model & 1/3, 1, \emph{3} \\
$\qquad \gamma$ & Multiplicative factor for actual process noise & 0.025, \emph{0.05}, 0.075, 0.1 \\
\hline
\end{tabular}
 \end{center}
  \caption{Parameter values used in simulations (default values in italics)}\label{table:parameters}
 \end{table}

For each different set of parameters, 50 random trajectories were generated according to the specifications described in \cite{li2016particle}. The initial target vector was chosen from a Gaussian with mean equal to the true target vector, and covariance as a diagonal matrix with spatial entries equal to 100 and velocity entries equal to 0.0005 (as in \cite{li2016particle}). For each trajectory, there was one set of measurements on which each filter was ran once. Figure 3 shows position estimates and particles obtained by our filter (designated as the `TT filter' in the following discussion) during a typical simulation.

To compare average errors, we also used the optimal mass transfer (OMAT) metric\cite{schuhmacher2008consistent} which  matches estimated target positions with the actual target positions so as to minimize total distance error (Li and Coates also used this metric in \cite{li2016particle}, and it is included in their Matlab testbench). The  five different filters tried in the simulations were the   PFPF\_EDH and  PFPF\_LEDH filters designed by Li and Coates; , a standard bootstrap particle filter with 1 million points; and the TT filter without and with the nonlinear correction described in Section~\ref{sec:nonlinear}. The three previous filters used were the best-performing filters investigated by Li and Coates in  \cite{li2016particle}. 

For each parameter set, to obtain final OMAT values we averaged over 50 different multitarget tracks over 40 time steps. Simulations were run with  MATLAB R2020a on a 64-bit computer with Intel Core i5 dual cores at 2.60 GHz with 6.0 GB total virtual memory. 

\subsection{Filter performance comparisons}
\begin{figure}[htb]
	\begin{center}
		\includegraphics[width=4in]{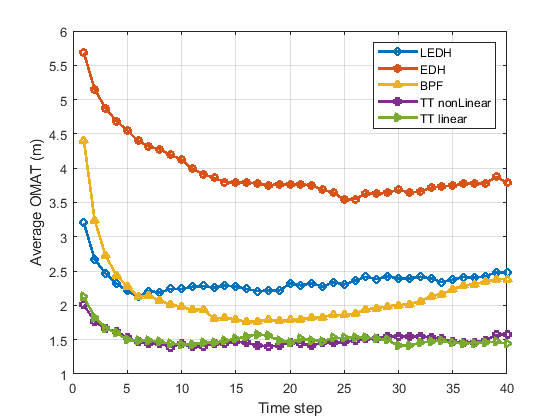}
			\caption{Average OMAT error per time step for $\sigma_w^2 = 0.1$ and other parameters at baseline}
		\label{fig:OMATperTime}
	\end{center}
\end{figure}

Figure \ref{fig:OMATperTime} shows the OMAT metric (averaged over 50 multitarget tracks) at each time step for the five tracking algorithms, for $\sigma_w^2 = 0.1$ and other parameters at baseline. 

Table \ref{table:baseline} compares the performance for the different algorithms for $\sigma_w^2 = 0.1$ and other parameters at baseline. The TT filter gives 36\% lower error than the PFPF\_LEDH, with less than 1/10 of the execution time. 

\begin{table}[hb]
\begin{center}
 \begin{tabular}{|m{2cm} | m{2.6cm} | m{3cm} | m{3cm}|} 
\hline
Filter& Particle Number & Average OMAT (m)& Execution Time per time step (s) \\
 \hhline{|=|=|=|=|}
TT linear & 144 & 1.515 & 0.14 \\
TT nonlinear &144 & {\bf 1.503} & 0.14 \\
PFPF\_LEDH & 500 & 2.352 & 1.78 \\
PFPF\_EDH & 500 & 3.962 & {\bf 0.02} \\
PFPF\_BPF & 1,000,000 & 2.123 & 6.52 \\
 \hline
\end{tabular}
 \end{center}
  \caption{Performance of different tracking filters with $\sigma_w^2=0.1$ and other parameters at baseline}\label{table:baseline}
 \end{table}

\begin{figure}[htb]
	\begin{center}
		\includegraphics[width=4in]{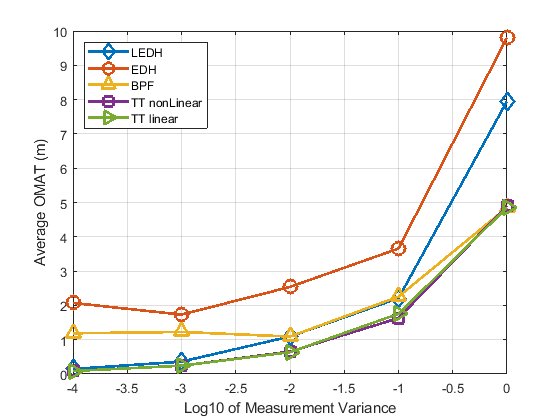}
			\caption{Average OMAT error per time step for different values of measurement variance ($\sigma_w^2$)}
		\label{fig:OMATmeasVar}
	\end{center}
\end{figure}

\begin{figure}[htb]
	\begin{center}
		\includegraphics[width=4in]{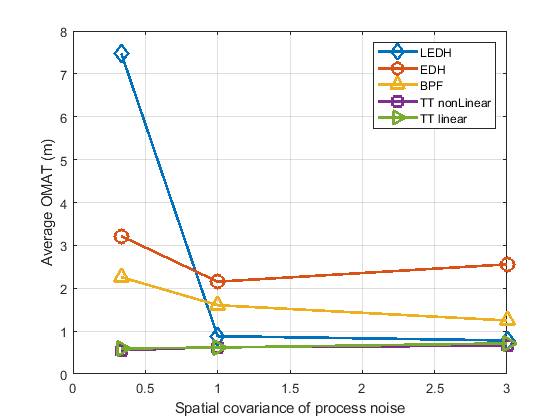}
			\caption{Average OMAT error per time step for different values of spatial covariance of filter’s process noise model ($\alpha$)}
		\label{fig:OMATspatialCov}
	\end{center}
\end{figure}

\begin{figure}[htb]
	\begin{center}
		\includegraphics[width=4in]{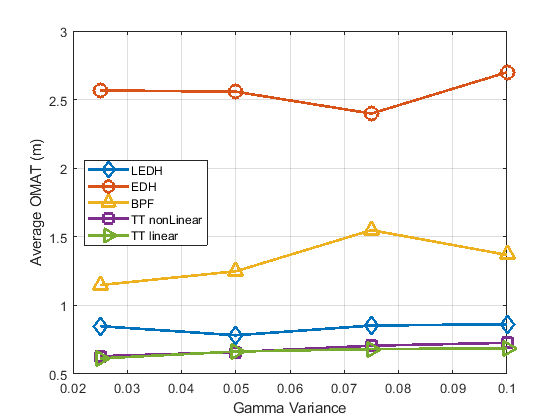}
			\caption{Average OMAT error per time step for different values of multiplicative factor for actual process noise ($\gamma$)}
		\label{fig:OMATgammaVar}
	\end{center}
\end{figure}

Figures \ref{fig:OMATmeasVar}-\ref{fig:OMATgammaVar}  compare the accuracy of the different algorithms for 
ranges of values for $\sigma_w^2$, measurement variance, filter's estimate of spatial covariance,  and process noise covariance respectively. The TT algorithms consistently gave the smallest error across all parameter values. Figures~\ref{fig:OMATspatialCov} and \ref{fig:OMATgammaVar}  which shows that only the TT algorithm shows consistent improvements when the filter's spatial covariance estimate or the process noise spatial covariance is decreased.   

\FloatBarrier

\subsection{Variants of the TT filter}

\begin{figure}[htb]
	\begin{center}
		\includegraphics[width=4in]{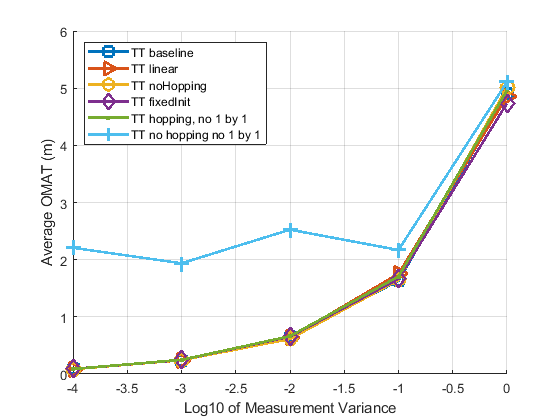}
			\caption{Average OMAT error per time step for different values of measurement variance}
		\label{fig:OMATmeasVarTT}
	\end{center}
\end{figure}

\begin{figure}[htb]
	\begin{center}
		\includegraphics[width=4in]{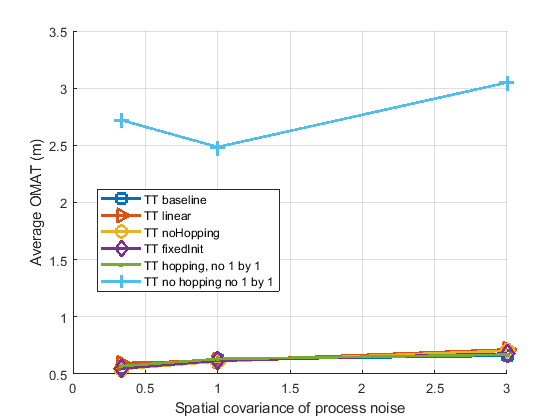}
			\caption{Average OMAT error per time step for different values of measurement variance}
		\label{fig:OMATspatialCovTT}
	\end{center}
\end{figure}

\begin{figure}[htb]
	\begin{center}
		\includegraphics[width=4in]{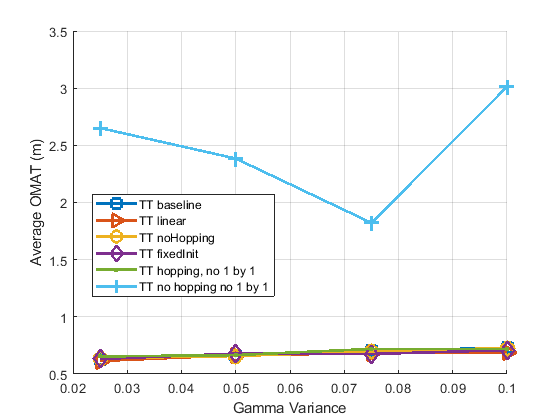}
			\caption{Average OMAT error per time step for different values of measurement variance}
		\label{fig:OMATgammaVarTT}
	\end{center}
\end{figure}

Figures~\ref{fig:OMATmeasVarTT}-\ref{fig:OMATgammaVarTT} compare the performance of several variants of the TT algorithm. The following variants are included:

\begin{itemize}
\item
TT baseline:  TT algorithm with the nonlinear correction described in Section~\ref{sec:nonlinear}, the hopping target correction described in Section~\ref{sec:hopping}, and the 1 by 1 correction method described in Section~\ref{sec:1by1}. This option also uses the randomized initialization used by Li and Coates in \cite{li2016particle}.
\item
TT linear: same as previous, but without the nonlinear correction described in Section~\ref{sec:nonlinear};
\item
TT noHopping:   same as TT nonLinear, except without the hopping target correction described in Section~\ref{sec:hopping};
\item
TT fixedInit: TT algorithm without any prior estimate of the target position vector at $t=0$. Targets were initialized to lie within 5 m of the central sensor, and the ML702LE estimation procedure described in Section~\ref{sec:NLLmeas} was used with the first measurement to obtain the target vector at the first time step $t=1$.
\item
TT hopping no 1 by1:  Same as TT nonLinear above, except without the 1 by 1 correction method described in Section~\ref{sec:1by1};
\item
TT no hopping no 1 by1:   Same as TT nonLinear above, except without neither the 1 by 1 correction method described in Section~\ref{sec:1by1} nor the hopping target correction described in Section~\ref{sec:hopping};
\end{itemize}

The figures show that  all variants perform nearly the same except when none of the corrections described in Section~\ref{sec:consistency} are applied.  We conclude that either correction method performs equally well, and there is no need for secondary correction.  We note also that the filters of Li and Coates require an initial guess target vector that is not too far from the true target vector to achieve convergence, while the TT filter converges as long as targets are placed near the center of the sensing grid.  This represents another distinct advantage of the TT filter.

\section{Conclusion}
We have documented the design,validation, and verification of a higher-performing, lower complexity multi-target tracking algorithm for a sensor network. Our TT filter achieves over 30\% reduction in error over the most accurate filter described in \cite{li2016particle} (as measured by the OMAT metric) while reducing the complexity by over 90\%. The TT algorithm also works equally well in scenarios where there is no prior knowledge of targets before tracking begins. 

For future work, we may examine larger systems with more targets and sensors.  We will also improve the uncertainty estimation for targets. Currently, only the covariance of the targets is computed at time $t$.  Since the target position distribution is not Gaussian, this is not an accurate indication of the target's uncertainty. By making use of nonlinear transformations, we hope to improve the uncertainty region.

\section*{Ackowledgements}
This research was supported by the Air Force Research Laboratory, and has been approved for public release: distribution unlimited (Case Number: 88ABW-2020-2798).

\FloatBarrier

\bibliographystyle{plain}
\bibliography{BiblioParticleFilter}
%
%\begin{thebibliography}{00}
%\bibitem{b1} G. Eason, B. Noble, and I. N. Sneddon, ``On certain integrals of Lipschitz-Hankel type involving products of Bessel functions,'' Phil. Trans. Roy. Soc. London, vol. A247, pp. 529--551, April 1955.
%\bibitem{b2} J. Clerk Maxwell, A Treatise on Electricity and Magnetism, 3rd ed., vol. 2. Oxford: Clarendon, 1892, pp.68--73.
%\bibitem{b3} I. S. Jacobs and C. P. Bean, ``Fine particles, thin films and exchange anisotropy,'' in Magnetism, vol. III, G. T. Rado and H. Suhl, Eds. New York: Academic, 1963, pp. 271--350.
%\bibitem{b4} K. Elissa, ``Title of paper if known,'' unpublished.
%\bibitem{b5} R. Nicole, ``Title of paper with only first word capitalized,'' J. Name Stand. Abbrev., in press.
%\bibitem{b6} Y. Yorozu, M. Hirano, K. Oka, and Y. Tagawa, ``Electron spectroscopy studies on magneto-optical media and plastic substrate interface,'' IEEE Transl. J. Magn. Japan, vol. 2, pp. 740--741, August 1987 [Digests 9th Annual Conf. Magnetics Japan, p. 301, 1982].
%\bibitem{b7} M. Young, The Technical Writer's Handbook. Mill Valley, CA: University Science, 1989.
%\end{thebibliography}

\end{document}